\documentclass[prb,11pt]{revtex4-2}

\usepackage{afterpage,placeins}
\usepackage{graphicx,mathtools,multirow,hyperref}
\usepackage{booktabs}
\usepackage{amssymb}
\usepackage{placeins}
\usepackage{color}

\begin{document}

\title{Visualization of the Challenges and Limitations of the Long-Term Sunspot Number Record}

\author{Andr\'es Mu\~noz-Jaramillo}
\affiliation{SouthWest Research Institute, Boulder, CO 80302, USA}
\affiliation{High Altitude Observatory, Boulder, CO 80301, USA}
\affiliation{National Solar Observatory, Boulder, CO 80303, USA}
\affiliation{\href{mailto:amunozj@boulder.swri.edu}{amunozj@boulder.swri.edu}}

\author{Jos\'e M.\ Vaquero}
\affiliation{Universidad de Extremadura, Departamento de F\'isica, M\'erida, Badajoz, Spain}
\affiliation{Universidad de Extremadura, Instituto Universitario de Investigación del Agua, Cambio Clim\'atico y Sostenibilidad (IACYS), Badajoz, Spain}

%



\begin{abstract}
The solar cycle periodically reshapes the magnetic structure and radiative output of the Sun and determines its impact on the Heliosphere roughly every eleven years.  Besides \textbf{this} main periodicity, it shows century long-term variations (including periods of abnormally low solar activity called grand minima). The Maunder Minimum (MM; 1645-1715) has generated significant interest as the archetype of a “Grand Minimum” in magnetic activity for the Sun and other Stars,suggesting a potential link between the Sun and changes in terrestrial climate.  Recent reanalyses of sunspot observations have yielded a conflicted view on the evolution of solar activity during the last 400 years (a steady increase vs. a constant level).  This has ignited a concerted community-wide effort to understand the depth of the MM and the subsequent secular evolution of solar activity.  The goal of this perspective is to review recent work that uses historical data to estimate long-term solar variability, and to provide context to users of these estimates that may not be aware of their limitations.  We propose a clear visual guide than can be used to easily assess observational coverage for different periods, as well as the level of disagreement between currently proposed sunspot group number series.
\end{abstract}

\flushbottom
\maketitle
%
%
\thispagestyle{empty}

\vspace{-2em}

\section*{Introduction}

The solar cycle was first hypothesized by Christian Horrebow in the 1770s and later rediscovered by Heinrich Schwabe in 1844, as a clear, roughly decadal modulation of the number of sunspots visible on the solar disk (see Figure \ref{fig:groups}).  Schwabe's results, publicized by Alexander von Humboldt,  motivated Rudolf Wolf, at the Zurich Observatory, to collect all prior available sunspot observations and use them to make the first determination of the wolf sunspot number (WSN) series -- a weighted average of the amount of sunspots and sunspot groups -- in 1848\cite{vaquero-vazquez2009}. From the beginning,  piecing together sunspot observations of different observers has proved to be a challenging endeavor --  an active field of research even today.  However, most users of sunspot series data are unaware of the nuances and challenges of piecing together a homogeneous series out of an heterogeneous set of observers and thus take the sunspot series for granted. The goal of this perspective is not to endorse a particular view of the secular evolution of solar activity over the last 400 years, but to provide a clear visual guide that users of historic sunspot data, as well as their derivate data products, can use to understand the limitations of historic sunspot observations.  This perspective is part of a concerted community-wide effort to raise awareness of this issues and provide users with a new, live, sunspot series that is more transparent and open to further improvement in the future. As such, we represent here a snapshot assessment current to 2018, that will be revised in the future as more historic data is unearthed and our methods improve leading to better reconstructions of long-term solar activity.

\subsection*{What Kinds of Studies Need to Be Careful With Their Use of Solar Long-Term Variability Data?}

There is a wide variety of studies that hinge on an accurate assessment of the evolution of solar activity.  Without attempting to perform a comprehensive review of all of them, it is important for users of the sunspot number series, as well as users of secondary products derived from their analysis, to be aware of the limitations of working with historical sunspot records.  These studies include, but are not limited to:
\begin{enumerate}
\itemsep-0.4em
\item Using the sunspot number series to reconstruct total and spectral solar irradiance \cite{Kopp-etal2016}, and other related indices.
\item Understanding the connection between solar activity and terrestrial climate change\cite{gray-etal2010}.
\item Using the Sun as a reference measuring stick to characterize stellar magnetic cycles \cite{wright2004}.
\item Using the MM as an illustrative scenario of what may happen in the future \cite{maycock-etal2015}.
\end{enumerate}
In particular, we would like to warn users of sunspot data to be wary of building highly speculative scenarios -- as these scenarios amplify the challenges and uncertainties of contextualizing the Sun as a star, and the role that solar variability plays in terrestrial climate change.

\subsection*{The Challenges of Assembling a Sunspot Number series}

The study of the secular evolution of solar activity during the last 400 years largely focuses on underpinning the transition between two extrema: on the one hand we have the Maunder Minimum (MM) -- a period where solar activity was abnormally low during the second part of the 17th century\cite{eddy1976}; on the other hand we have the modern maximum (including the strongest cycles for which we have direct observations) taking place during the second half of the 1900's\cite{usoskin-etal2003}.

We can confidently determine solar activity levels since the beginning of the 20th century, and the majority of current studies that use information from documentary sources suggest very low levels solar activity during the MM\cite{usoskin-etal2015,carrasco-etal2015,vaquero-etal2015,vaquero-etal2016}, which is confirmed by cosmogenic isotopes\cite{asvestari-etal2017}.  However, it is difficult to ascertain activity levels in between.  The main challenge arises from having to piece together heterogeneous observers with:
\begin{enumerate}
\itemsep-0.4em
\item Differences in the telescope aperture, its optical quality, and observational method used by different observers.
\item Differences in each observer's visual acuity (as well as possible degradation with age).
\item Differences in each observer's definition of a sunspot group and how they should be counted.
\item Lack of uniform observational coverage and consistency.
\item Determining which observer should be used as the standard against which all other observers are calibrated (absolute calibration).
\end{enumerate}

\begin{figure}[b!]
\centering
\includegraphics[width=\linewidth]{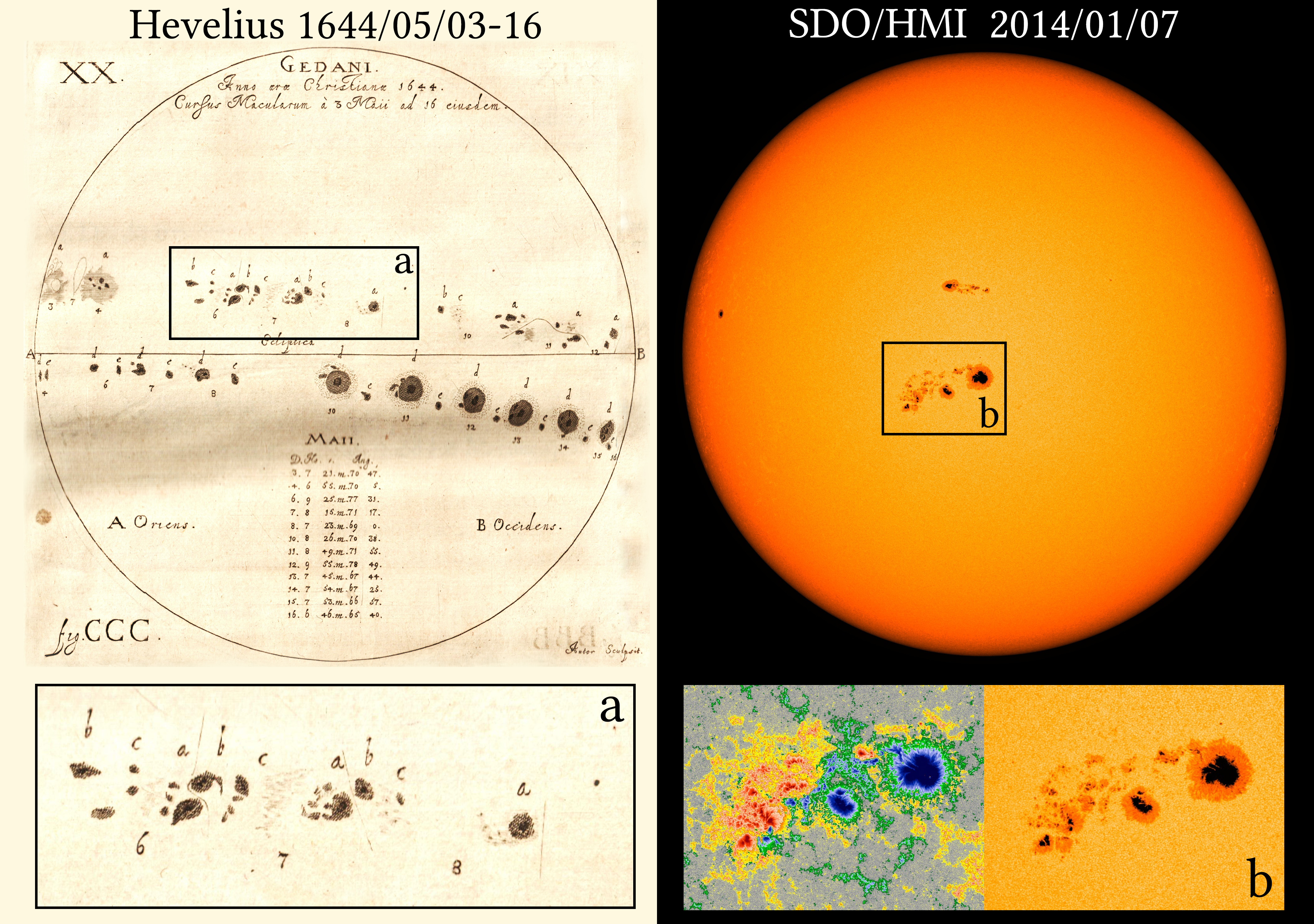}
\caption{Historical (left) vs. modern (right) observations of sunspot groups.  Historical observers often used the same disk to draw the time evolution of sunspot groups as can be seen in the close up (a) where Hevelius recorded the evolution of three different sunspot grous (a,b,c) on the 6th, 7th, \& 8th of May-1644 (faint lines indicate when a day ends and the next begins). Even with measurements of the magnetic field, shown in the close-up panel using green-blue (red-yellow) to indicate positive (negative) magnetic fields, it is challenging to tell different sunspot groups apart when they are close to each other.  Both close-up panels share the same scale so visible structures across this figure are directly comparable.  Left panel courtesy of the Library of the Astronomical Observatory of the Spanish Navy.  Right panel courtesy of NASA's Solar Dynamics Observatory.}
\label{fig:groups}
\end{figure}

\noindent These challenges are compounded when working with early historical data prior to the mid-19th century (see Figure \ref{fig:groups}), due to the need for dealing with:
\begin{enumerate}
\itemsep-0.4em
\item Significant observational gaps.
\item Latin originals.
\item Observers that did not have a common vocabulary to describe the same phenomena.
\item Surveys with scientific objectives unrelated to measuring solar activity.
\item Drawings that combine multiple days of observation on the same solar disk and/or arbitrarily scale objects to make ``better" use of the empty space of the quiet Sun.
\end{enumerate}

\subsection*{Weak or Strong Secular Evolution? The Wolf Sunspot Number vs.\ the Group Sunspot Number}

Our understanding of long term variability for more than a hundred years has been based on the WSN.   This changed in 1998 when Hoyt \& Schatten's \cite{hoyt-schatten1998} revisited and collected all available historic data in 1998 and used them to define a new sunspot-based series: the group sunspot number (GSN). As opposed to Wolf's number, which uses a weighted average of the amount of visible sunspots and sunspot groups (10$\cdot$No.\ Groups + No. Sunspots), the GSN is based exclusively on the number of groups.

The GSN painted a different picture of the secular evolution of solar activity during the last 400 years than the WSN (see solid pink and black dashed lines in Figure \ref{fig:data}(a): one in which the 1800's and 1700's display levels of activity that are nearly half of those during the 1900's; leading to the idea of the modern period being as unusual in its strength as the MM was in its weakness\cite{usoskin-etal2003}.  However, recent work\cite{clette-etal2014} has called into question the methods used to calculate \textit{both} the WSN and the GSN.  One of the important results of the subsequent conversation is that we are no longer certain that the modern period is truly extraordinary (i.e. activity levels after the MM may have been comparable to those of the 20th century) so care must be taken when using the old GSN as basis for modern exceptionalism.

\subsection*{From the Maunder Minimum to the Present: As Much Detective Work as Science}

While both the GSN and WSN are important for our understanding of long-term solar variability, in this perspective we focus exclusively on GSN reconstructions.  The reason is that, at the moment, the GSN series is the only one for which we have the original raw data, making it the only series that can potentially be extended all the way back to the 17th century.  However, there is currently a parallel effort to improve the WSN as well\cite{clette-lefevre2016}.

\begin{figure}[t!]
\centering
\includegraphics[width=\linewidth]{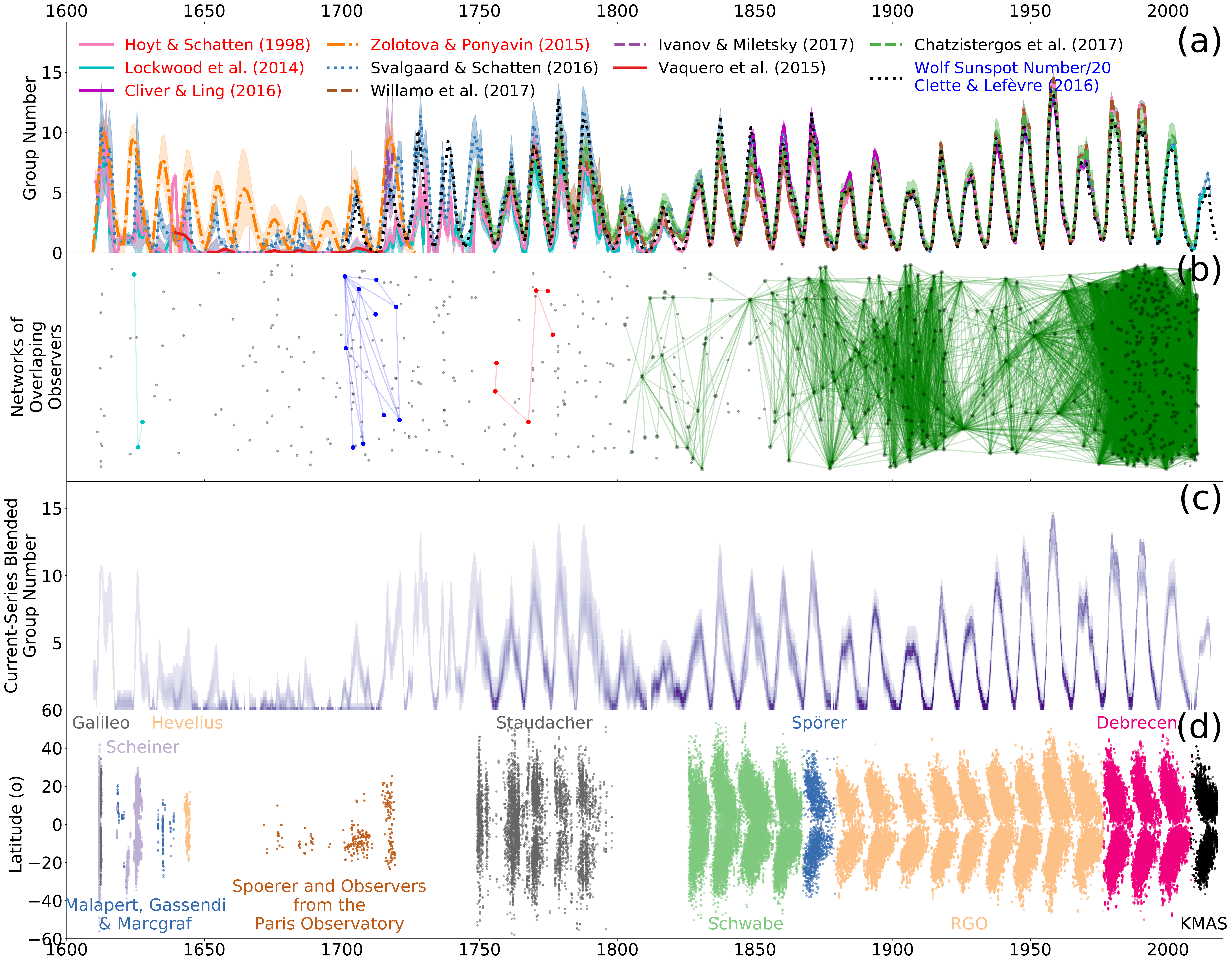}
\caption{\textbf{Long-term evolution of solar activity.}  There are currently a wide variety of different reconstructions of solar activity using different techniques (a). Series whose name is highlighted in black\cite{vaquero-etal2015,svalgaard-schatten2016,willamo-etal2017,chatzistergos-etal2017,ivanov-miletsky2017} are considered the best estimates we currently have and they are used in our dispersion analysis.  Series whose name is highlighted in red are included for completeness, but are considered to be superseded\cite{hoyt-schatten1998}, provisional \cite{lockwood-etal2014,cliver-ling2016}, or refuted\cite{zolotova-ponyavin2015} by more recent work. The International (Wolf) Sunspot Number divided by 20\cite{clette-lefevre2016}, in blue, is included for reference. All series (using different colors and line styles) are shown as provided with no effort to scale them to each other. The challenge of establishing a connection between historic and modern observers is visualized by constructing a graph network where each node is a unique observer and each edge connects observers with sufficient observational overlap (b).  Only observers within a colored network have significant overlap with each other. We propose the combination of current series\cite{vaquero-etal2015,svalgaard-schatten2016,willamo-etal2017,chatzistergos-etal2017,ivanov-miletsky2017} into a single heat map (c), instead of picking and choosing only one of them, as a more conservative way of assessing long-term solar activity.  We also include the long-term butterfly diagram (d), using the latitude and time observed for individually observed groups, as valuable supplemental data to the sunspot series.  We include all published modernized latitude-time sunspot group observations (using different colors to indicate different observers and surveys)\cite{ribes-nesmeribes1993,arlt2009,vaquero-etal2011,senthamizhpavai-etal2015,diercke-arlt-denker2015,vaquero-etal2015b,baranyi-etal2016,arlt-etal2016,gyori-etal2017,tlatov-etal2017,vokhmyanin-zolotova2018}, and preliminary data based on original drawings, to provide a complete picture of long-term solar variability.}
\label{fig:data}
\end{figure}

\textbf{Two main approaches have been proposed for piecing together GSN observations from 1750 to the present:}
\begin{enumerate}
\itemsep-0.4em
\item The use of a carefully selected skeleton of main observers that acts as a scaffold to combine all available observers into a single series \cite{svalgaard-schatten2016,cliver-ling2016,chatzistergos-etal2017}.
\item Using the fraction of active (at least one sunspot visible on disk) vs.\ quiet (no sunspots visible) days to determine the visual acuity of historical observers enabling their calibration to a highly detailed ``objective-observer'' like the Royal Greenwich Observatory photographic series\cite{usoskin-etal2016,willamo-etal2017}.
\end{enumerate}
Prior to 1750 things become more complicated due to the nature, quality, and coverage of historical observations.  There are currently four main proposed approaches (the first of which is only included for completeness):
\begin{enumerate}
\itemsep-0.4em
\item The construction of highly speculative scenarios based on the assumption that: a. 11 year solar cycle operated continuously during the MM. b. observations during the MM are flawed because observers were not aware of the existence of the solar cycle nor interested in its associated phenomena\cite{zolotova-ponyavin2015}.  These speculative scenarios have been since refuted\cite{usoskin-etal2015} and should not be used as estimates of solar activity for the MM.
\item The estimation of activity based on the fraction of active vs.\ quiet days by using direct comparison with the relationship between the active day fraction and activity of the period between 1850 and 1995\cite{kovaltsov-etal2004,vaquero-etal2015}.
\item The use of the highest daily count of sunspot groups for each year by any observer during the MM (``the brightest star”) under the assumption that such occurrences of spottedness represented relatively rare episodes of activity that were more likely to be noticed and reported.\cite{svalgaard-schatten2016}.
\item The estimation of activity level based on the width of active latitudes (i.e.\ butterfly wings)\cite{ivanov-miletsky2017}.
\end{enumerate}
All of these proposed series, along with the the WSN\cite{sidc,clette-lefevre2016}, the (now) superseded series of Hoyt and Schatten\cite{hoyt-schatten1998}, and the provisional series of Lockwood et al.\cite{lockwood-etal2014}, and Cliver \& Ling\cite{cliver-ling2016}, are illustrated in Figure \ref{fig:data}(a).  They form part of an ongoing community effort to identify the best methods to ascertain the level of solar activity during the last four centuries.


\section*{The First Problem: The Sunspot Series is Taken for Granted}

One of the biggest problems leading to unsubstantiated claims and conclusions is that the sunspot series (and automatically any derived studies) is largely perceived as a consistent, completed, and accurate record.  This perception is so commonplace that the sunspot record is currently listed by the Guinness World Records as the ``longest continuous observational science data"\cite{guinness-records}.  

Figure \ref{fig:data}(b) clearly illustrates the challenge of trying to estimate the relative amplitude of historic vs.\ modern solar activity. It shows a network graph of all the 715 observers (from Galileo until the present) for which we currently have sunspot number observations\cite{vaquero-etal2016}.  Each node in the network denotes a unique observer temporally centered at the middle of its observational period.  Two observers are connected if they:


\begin{enumerate}
\itemsep-0.4em
\item Have at least 60 different overlapping days of observations (each unique pair of days within 2 days of each other).
\item Each observer reported at least 4 different values of group numbers during their observational survey (a bare minimum to enable cross-calibration).
\end{enumerate}
The main clusters of connected observers are highlighted using different colors, demonstrating the difficulty of establishing a direct observational connection between the present and years prior to 1850.

\subsection*{Solution: Use Series Disagreement as Empirical Evidence of Uncertainty}

Instead of picking and choosing a particular series for their research, we suggest for users to see existing proposed series as a map of the inherent empirical uncertainty of such reconstruction.   We illustrate this idea in Figure \ref{fig:data}(c) by combining, year-by-year, all available current series\cite{vaquero-etal2015,svalgaard-schatten2016,willamo-etal2017,chatzistergos-etal2017,ivanov-miletsky2017} using the reported activity level and uncertainty of each series.  For each year between 1610 and 2017 we:

\begin{enumerate}
\itemsep-0.4em
\item Define a normalized Gaussian probability density function (PDF) separately for each series with valid data for that year.  These PDFs are centered on each series' reported activity level, and have a standard deviation equal to each series' reported uncertainty.  This procedure converts the reported activity level for each year and each series into a continuum of possibilities that can be added.
\item Superimpose the PDFs of all series with valid data in that year using simple addition.
\item Evaluate the combined/added PDFs using a range of group number values (0-20) and append the resulting array into a 2D array that is used for visualization.  In this array each row denotes a group number, each column a year, and each cell contains the combined/added PDF values.
\end{enumerate}

The resulting array can be used to visualize a ``heatmap" that darkens where the series agree, or lightens where they disagree.  Normalized PDFs naturally highlights regions where there is more than one proposed series, and dims regions that have not been analyzed by multiple groups.  Figure \ref{fig:data}(c) provides a more conservative overview of the possible ranges of solar activity over the past 400 years (as well as its uncertainty) than any single one of the proposed series shown in Figure \ref{fig:data}(a).  It is important, however, to stress that this proposed approach is only temporary and will be superseded by the revised framework and protocols that the solar community is currently working on.




\section*{The Second Problem: Users are Unaware of the Prevalence of Observational Gaps}

The traditional way of visualizing solar long-term variability data (see Figures \ref{fig:data}(a) and \ref{fig:mm_vs_modern}(a)) is failing users by not making a strong distinction between the quality and abundance of observations in the modern vs.\ historical periods.  This absence is particularly severe and lacking in conventional butterfly diagrams (see Figure \ref{fig:data}(d)), which use the latitude and time of observed sunspot groups to illustrate the equator-ward migration of active latitudes with the progression of the cycle.

The marked difference in abundance of observations between modern and historical periods is visible in Figure \ref{fig:mm_vs_modern}(b), showing a grid of daily observations stacked in 30 day intervals.  Each grid square corresponds to a unique day and is colored to show whether it was active (purple; at least one observer reported a sunspot group), quiet (blue; all observers reported a naked Sun), or missing (white; no observations available).  A quick comparison between the current available GSN series during the MM\cite{vaquero-etal2015,svalgaard-schatten2016,ivanov-miletsky2017} and our observational coverage (Figure \ref{fig:mm_vs_modern}(c)) highlights the speculative nature of our understanding of historic solar activity (especially during the first half of the MM).

\section*{Solution: Masking Solar Variability Data According to their Observational Coverage}

\begin{figure}[t!]
\centering
\includegraphics[width=\linewidth]{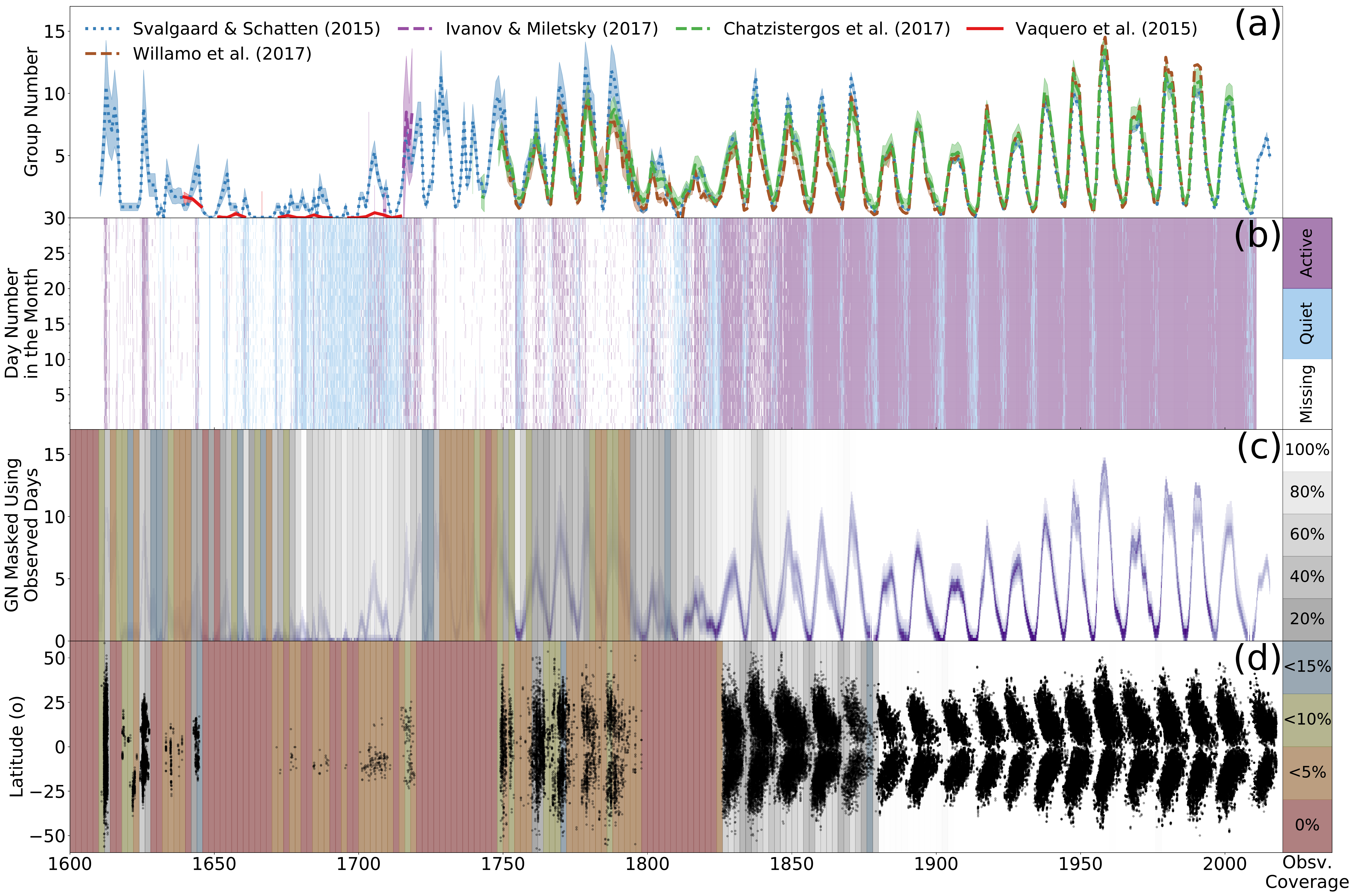}
\caption{\textbf{Observational coverage of historical sunspot group data.} Traditional visualization of sunspot number series use continuous curves and shaded areas to indicate solar activity level and its uncertainty (a).  However, they fail to illustrate how many observations were used to estimate activity level. We use a grid of daily observations stacked in 30 day intervals where each grid square is a day colored to show whether it was active (purple; at least one observer reported a sunspot group), quiet (blue; all observers reported a naked Sun), or missing (white; no observations available) to illustrate the stark difference in coverage between the MM and modern period (b). We propose to use color and transparency to mask the proposed series such that highly speculative intervals are no longer visible (c) and thus the user's perception of the historical period differs significantly than that of the modern period.  In this color-transparency scheme, white (red) represents 100\% (0\%) observational coverage, while other colors represent intervals or coverage values (i.e.\ orange indicates coverage between 0\%-5\%, yellow between 5\%-10\%, etc.).  Above 15\% we use a continuous colormap going from gray to white.  Observational coverage is particularly poor for the historic butterfly diagram data (d).  Because of this, we plot the color mask behind the data or nothing would be visible.  Comparing historic and modern butterfly diagram data must be done with extreme care.}
\label{fig:mm_vs_modern}
\end{figure}

To address this deficiency we propose to mask the sunspot series (and butterfly diagram) encoding observational coverage using color and transparency (the more sparse the observational coverage the more opaque the mask).  The calculation and over-plotting of a biennial mask of observational coverage (Figure \ref{fig:mm_vs_modern}(c)) shows how different coverage is during the modern vs.\ historic periods:  The modern period is nice and white, fully covered and clearly distinct (as it has been shown traditionally), while the historic period has many intervals that are highly opaque and colored, masking highly speculative intervals, and warning users to be careful with how they use historic data and frame their conclusions.

For the butterfly diagram (Figure \ref{fig:mm_vs_modern}(d)) coverage is so sparse (typically less than 5\%), than it's better to draw the mask behind the data (or nothing would be visible).  However, by including this information in the visualization it is quite apparent to users how challenging it is to compare historic and modern butterfly diagram observations in a straightforward manner as has been done since the seminal work of Ribes \& Nesme-Ribes in 1993 \cite{ribes-nesmeribes1993}.


\section*{A New Way of Visualizing Solar Long-term Variability Data}

\begin{figure}[t!]
\centering
\includegraphics[width=\linewidth]{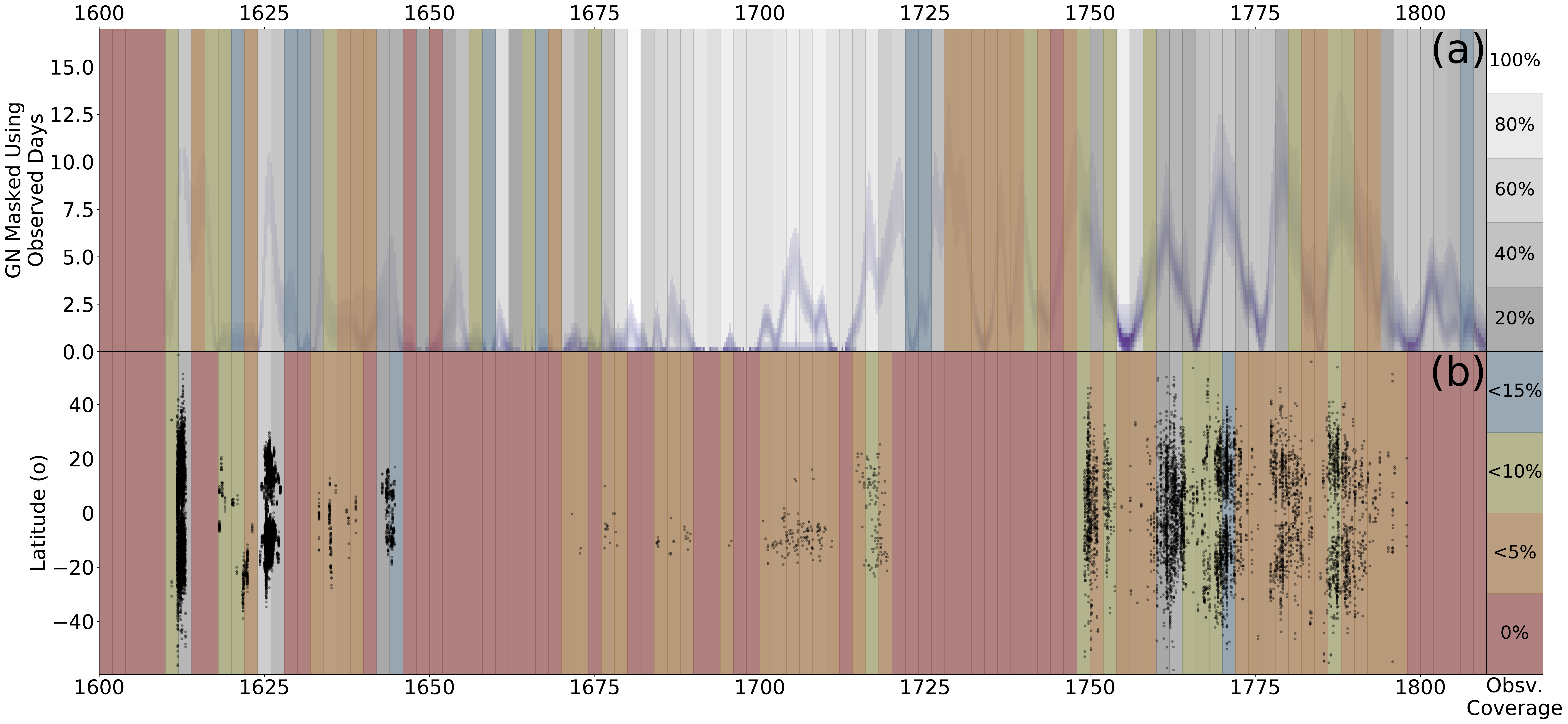}
\includegraphics[width=\linewidth]{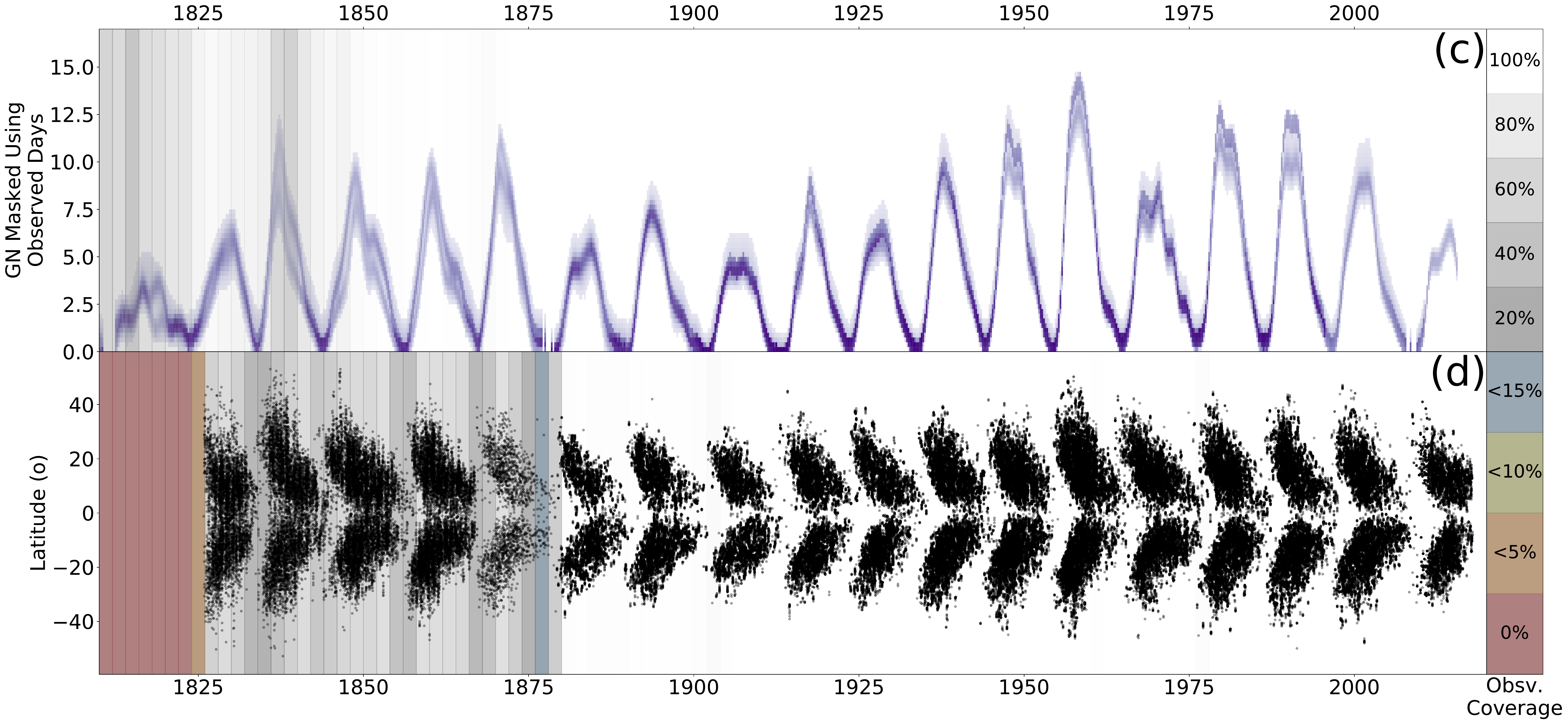}
\caption{\textbf{New way of visualizing observational gaps in historical solar data.} Instead of the traditional way of looking at the sunspot number series (Figs.~\ref{fig:data}(a) \& \ref{fig:mm_vs_modern}(a)), we propose a new way of visualizing historic data that captures simultaneously current understanding of long-term variability, our disagreements regarding the level of solar activity observed during the last 400 years, and information on the quality of observational coverage for different historical periods (a \& c).  For completeness we also include all currently available butterfly diagram data as well as its observational coverage (b \& d).}
\label{fig:coverage}
\end{figure}

Our objective is to propose a new, improved, way of visualizing the evolution of long-term solar variability (see Figures \ref{fig:mm_vs_modern} \& \ref{fig:coverage}) that permits scientists of any field to understand the limitations of our data and the highly speculative nature of our understanding of solar variability in the 1600s and 1700s.  We achieve this by combining our best assessments of the evolution of solar activity during the last 400 years, our disagreements regarding the best methods to make such assessment, and the availability of historical observations into clear visual cues.

Breaking the last four centuries into two intervals of 200 years (see Figure \ref{fig:coverage}), clearly highlights how much better solar observations became since Heinrich Schwabe started the seminal observational survey that led to the discovery of the solar cycle.  It also highlights how one of the biggest challenges of working with data from the 1600s and 1700s is the presence of long periods with poor observational coverage.

\section*{A Road-map into the Future}

The results presented in this perspective need to be taken as a snapshot that is likely to change as more data are discovered and our methods improve.  In particular, our community still needs to take advantage of recent advances in calibration techniques\cite{dudokdeWit2011}, better methodologies for quantifying uncertainties, recovery of still unexploited historical documents\cite{hayakawa-etal2018}, re-analysis of original sunspot drawings\cite{diercke-arlt-denker2015}, and further comparison to other proxies of solar activity\cite{svalgaard2016,asvestari-etal2017}, to maximize informational synergy.  There is currently an ongoing concerted effort to combine the strengths of all proposed methods to address their weaknesses. The end goal is an community-vetted and accepted version, with provision made for further updates as more data is unearthed and our understanding increases.  We expect that a new, better, open-source, official sunspot number series will be released in early 2020 by the SILSO World Data Center.


We are optimistic that the coming decade will see significant improvements in our assessment of long-term solar variability and its associated uncertainty.  However, no amount of sophisticated algorithms will be a substitute for the identification and proper translation of historical documents, as well as for interdisciplinary teams involving astrophysicists, historians of science, and philologists to revisit historical observations of sunspots\cite{carrasco-etal2015,neuhauser-neuhauser2016,hayakawa-etal2018}.  New data and deeper knowledge of the way historical observers worked are of critical importance for improving our understanding of the earlier period of solar observations including the Maunder Minimum.

\sloppy
\section*{Software Repository}

A Python Jupyter notebook, along with the necessary data to reproduce all plots presented here can be found at: \href{https://github.com/amunozj/NatAs\_SN\_Perspective}{https://github.com/amunozj/NatAs\_SN\_Perspective}.  We request that you use data in this repository only to reproduce our analysis, but get them from the original sources if you intend to use them for research purposes.

\section*{Acknowledgements}
We thank two anonymous referees and the editorial board for valuable feedback and suggestions.   We also thank the International Space Science Institute (ISSI-Bern) and the members of its team for the \href{http://www.issibern.ch/teams/sunspotnoser/}{Recalibration of the Sunspot Number Series} for providing the support and insight that made this perspective possible.  We also thank the ISSI team and Leif Svalgard for useful comments and suggestions.

This research is partly funded by the NASA Grand Challenge grant NNX14AO83G and NASA LWS grant NNX16AB77G, by the Economy and Infrastructure Counselling of the Junta of Extremadura through project IB16127 and grant GR15137 (co-financed by the European Regional Development Fund), and by the Ministerio de Economía y Competitividad of the Spanish Government (AYA2014-57556-P and CGL2017-87917-P).

Sunspot number series data are provided by the World Data Center SILSO, Royal Observatory of Belgium, Brussels, or kindly provided by the original authors\cite{zolotova-ponyavin2015,vaquero-etal2015,chatzistergos-etal2017,ivanov-miletsky2017}.  Butterfly diagram data are provided by the St.\ Petersburg State University\cite{vokhmyanin-zolotova2018}, the Historical Archive of Sunspot Observations (HASO) of the Universidad de Extremadura\cite{vaquero-etal2011,vaquero-etal2015b}, the Leibniz Institute for Astrophysics Potsdam (AIP)\cite{arlt2009,senthamizhpavai-etal2015,diercke-arlt-denker2015,arlt-etal2016}, the Debrecen Observatory\cite{baranyi-etal2016,gyori-etal2017}, and the Kislovodsk Astronomical Mountain Station\cite{tlatov-etal2017}.

\section*{Author contributions statement}

J.M.V. initiated the idea and collaboration behind the perspective. A.M.-J. condensed the message in visual and written form. All authors wrote and reviewed the manuscript.

\section*{Additional information}

\subsection*{Competing interests}

The authors declare no competing interests.

\sloppy

\end{document}